\documentclass[a4paper]{jpconf}
\usepackage{graphicx}
\begin{document}
\title{Matrix-Product-State Algorithm for Finite Fractional Quantum Hall Systems}

\author{Zhao Liu and R. N. Bhatt}
\address{Department of Electrical Engineering, Princeton University, Princeton, New Jersey 08544, USA}
\ead{zhaol@princeton.edu}

\begin{abstract}
Exact diagonalization is a powerful tool to study fractional quantum Hall (FQH) systems. However, its capability is limited by the exponentially increasing computational cost. In order to overcome this difficulty, density-matrix-renormalization-group (DMRG) algorithms were developed for much larger system sizes. Very recently, it was realized that some model FQH states have exact matrix-product-state (MPS) representation. Motivated by this, here we report a MPS code, which is closely related to, but different from traditional DMRG language, for finite FQH systems on the cylinder geometry. By representing the many-body Hamiltonian as a matrix-product-operator (MPO) and using single-site update and density matrix correction, we show that our code can efficiently search the ground state of various FQH systems. We also compare the performance of our code with traditional DMRG. The possible generalization of our code to infinite FQH systems and other physical systems is also discussed.
\end{abstract}

\section{Introduction}
Topologically ordered phases of matter attract a great deal of interest currently in condensed matter physics. Fractional quantum Hall (FQH) states \cite{laughlin83,mr}, as one of the most well-known topological states, provide examples of some of the most exotic features such as excitations with a fraction of the electron charge that obey anyonic statistics \cite{fstat,fstatqh} and are essential resources in topological quantum computation \cite{topol-quantum-computing}.

Exact diagonalization (ED) in small systems has traditionally been the central numerical technique of theoretical research of FQH states. Despite its considerable success in studying some of the robust FQH states like the $\nu=1/3$ Laughlin state \cite{haldane-ed}, the largest system size that ED can reach is seriously limited by the exponential growth of the Hilbert space. This weakens its capability in understanding some more complex FQH systems, for example those with non-Abelian anyons and Landau level mixing. Therefore, new algorithms, like density-matrix-renormalization-group (DMRG) \cite{white}, were applied to FQH systems and the computable system size was increased by almost a factor of two compared with ED \cite{shibata,bk,feiguin,kovrizhin,jzhao,hu}.

Very recently, it was realized that some model FQH states and their quasihole excitations have exact matrix-product-state (MPS) representation \cite{zaletel,estienne,estienne2,yangle}. Motivated by this, and considering that the DMRG algorithm can be apparently formulated in the MPS language \cite{mps}, here we report a MPS code for finite FQH systems on the cylinder geometry, which is different from the one developed for infinite systems \cite{frank}. We will show the structure of our code, explain how to use it to search the FQH ground states, compare the performance of our code with ED and traditional two-site DMRG algorithm, and discuss possible generalizations.

\section{Code structure}
\begin{figure}
\includegraphics[width=16pc]{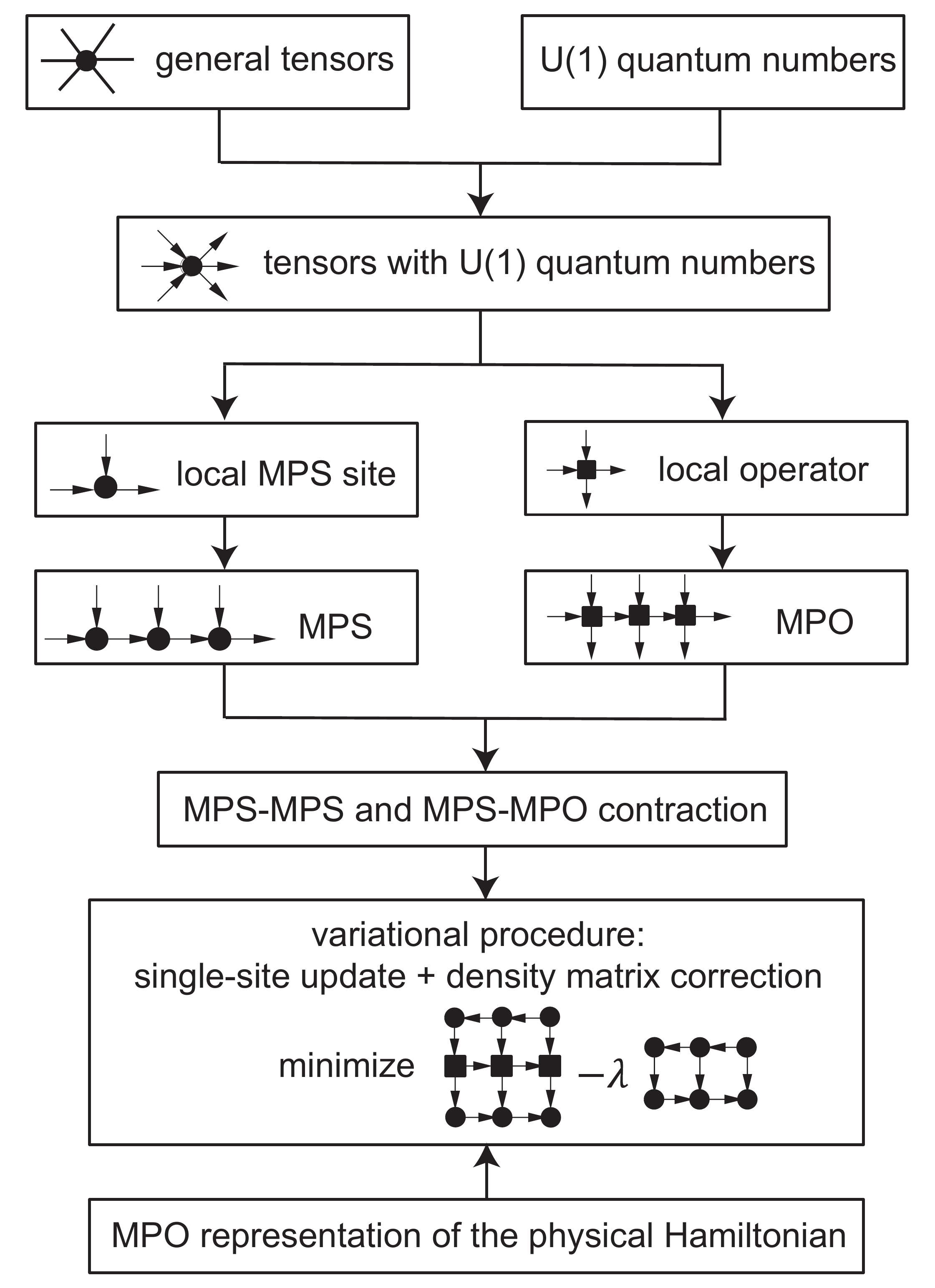}\hspace{3pc}%
\begin{minipage}[b]{16pc}\caption{\label{structure}The structure of our MPS code. The most basic part is located on the top, based on which we can implement the whole MPS algorithm step by step to search the ground state and ground energy. The arrows on the bonds of tensors represent $U(1)$ currents. The concepts of $U(1)$-symmetric tensors and their graphical representation are explained in Ref.~\cite{u1tensor}. The details of the variational procedure can be found in Ref.~\cite{mps}.}
\end{minipage}
\end{figure}

We show the structure of our code in Fig.~\ref{structure}. The most basic part is the implementation of general tensors and $U(1)$ quantum numbers. Based on that, we can construct tensors that conserve $U(1)$ quantum numbers \cite{u1tensor}. Each site in the MPS (MPO) is a special case of this kind of tensor with three (four) indices. Then we can implement the chains of these sites, namely MPS and MPO, and the contraction between them. Finally, with the MPO representation of the physical Hamiltonian as an input, we can do variational procedure (single-site update \cite{mps} and density matrix correction \cite{dmcorrect}) to minimize the expectation value of the energy $\lambda$.

\section{Solve the FQH problem}
\label{fqh}
For a FQH system with $N_e$ electrons and $N_s$ flux in a single Landau level on the cylinder of circumference $L$ (in units of the magnetic length), any
translational invariant two-body Hamiltonian can be written as
\begin{eqnarray}
H=\sum_l\sum_{m>0}\sum_{n>0} V_{m,n}c_{m+n+l}^\dagger c_{n+l}^\dagger c_{m+2n+l} c_l + h.c.
+\sum_l\sum_{m>0} V_{m,0}c_{m+l}^\dagger c_{l}^\dagger c_{m+l} c_l,
\label{hamil}
\end{eqnarray}
where $c_l^\dagger$ ($c_l$) creates (annihilates) an electron on the Landau level orbital $l=0,1,...,N_s-1$, and $m$ and $n$ determine
the range of $H$. The total
electron number $N_e=\sum_{l=0}^{N_s-1}c_l^\dagger c_l$ and total quasi-momentum $K=\sum_{l=0}^{N_s-1}lc_l^\dagger c_l$ are $U(1)$ good
quantum numbers.

%\begin{center}
%\begin{table}
%\centering
%\caption{\label{t1}We consider the matrix elements $V_{m,n}$ of the $\mathcal{V}_1$ pseudopotential at $L=20$. In the second column, we show the number of $V_{m,n}$ satisfying $|V_{m,n}|\geq 10^{-12}$ for each $n\leq12$. In the third column, we show $d_n$ (dimension of $A_n$, $B_n$ and $C_n$) in the exponential expansion approximation (with error $\varepsilon=10^{-6}$) of these $V_{m,n}$ for each $n\leq12$. We also give the total bond dimension of MPO obtained by two methods mentioned in Sec.~\ref{fqh}. If we keep all $V_{m,n}$ with $|V_{m,n}|\geq 10^{-12}$ and $m,n\leq\Lambda=12$, $D_{\textsc{MPO}}=348$. However, one can see that the exponential expansion approximation of $V_{m,n}$ indeed significantly reduces $D_{\textsc{MPO}}$.}
%\begin{tabular}{ccc}
%\br
%$n$&number of $V_{m,n}$ ($|V_{m,n}|\geq 10^{-12}$)&$d_n$\\
%\mr
%0&26&9\\
%1&25&8\\
%2&24&8\\
%3&23&7\\
%4&22&7\\
%5&21&6\\
%6&20&5\\
%7&18&5\\
%8&17&4\\
%9&16&3\\
%10&14&3\\
%11&13&2\\
%12&11&2\\
%\br
%$D_{\textsc{MPO}}$&348&179\\
%\br
%\end{tabular}
%\end{table}
%\end{center}

In our code, the procedure to search for the ground state of $H$ in a fixed $(N_e,K)$ sector is as follows:
\begin{itemize}
\item Construct an initial MPS state $|\Psi_0\rangle$ (either a random or a special state) with fixed $(N_e,K)$.
\item Construct a MPO representation of $H$, which can be generated by the finite state automaton \cite{machine}. There are two choices when applying the finite state automaton method. The simplest one is to keep all $V_{m,n}$ satifying $|V_{m,n}|$ larger than some threshold (for example $10^{-12}$) and $m,n\leq\Lambda$, where $\Lambda$ is the truncation of the interaction range. The bond dimension of the MPO, $D_{\textrm{MPO}}$, obtained in this way is proportional to $\Lambda^2$ \cite{frank}. The other choice is numerically much more efficient. For each $n\leq\Lambda$, we first select all $V_{m,n}$ satisfying $|V_{m,n}|$ larger than some threshold (for example $10^{-12}$), and then approximate them by an exponential expansion $C_n A_n^{m-1} B_n$, with an error $\varepsilon_{m,n}=|V_{m,n}-C_n A_n^{m-1} B_n|$. Here $C_n$, $A_n$ and $B_n$ are real $1\times d_n$, $d_n\times d_n$ and $d_n\times 1$ matrices, respectively, whose optimal values can be found by the state space representation method in the control theory \cite{private}. This exponential expansion of $V_{m,n}$ can reduce the bond dimension of MPO by a factor of two compared with the first choice and is especially suitable for long-range interactions \cite{private,vidal}. We find that typically $\Lambda\approx10$ and $\varepsilon\approx10^{-8}$ are enough for a good representation of $H$.
\item Do the variational procedure with the MPO representation of $H$ for the initial MPS state $|\Psi_0\rangle$. In the density matrix correction $\widetilde{\rho}=\rho+\alpha\Delta\rho$ \cite{dmcorrect}, $\alpha$ should be a small number (we choose $\alpha=10^{-5}$ in the following). Eigenstates of the corrected reduced density matrix $\widetilde{\rho}$ with eigenvalues larger than $\eta$ are kept. The smallest and largest allowed number of kept states can also be set. When the energy goes up, the density matrix correction is switched off. After the energy converges, we get a candidate of the ground state of $H$.
\item Try different values of $\alpha$ and $\eta$ to make sure we are not trapped in local energy minimum.
\end{itemize}

\section{Comparison with exact diagonalization}
In this section, we study a system of $N_e=12$ electrons at filling $\nu=1/3$ with $N_s=3N_e-2=34$ and $L=20$, which is easily accessible for ED. By comparing the MPS results with ED, we demonstrate that there are two error sources in our MPS code: one is the quality of the MPO representation of the Hamiltonian, the other is the density matrix truncation.

We use Haldane's $\mathcal{V}_1$ pseudopotential \cite{v1} as the Hamiltonian, for which
\begin{equation}
V_{m,n}=-\frac{16\pi^2\sqrt{2\pi}}{L^3}m(m+2n)e^{-\frac{2\pi^2}{L^2}[(m+n)^2+n^2]}.
\end{equation}
The ground state in the $(N_e,K=3N_e(N_e-1)/2)$ sector is the exact Laughlin state with exactly zero ground energy.
\begin{figure}
\centerline{\includegraphics[width=0.8\linewidth]{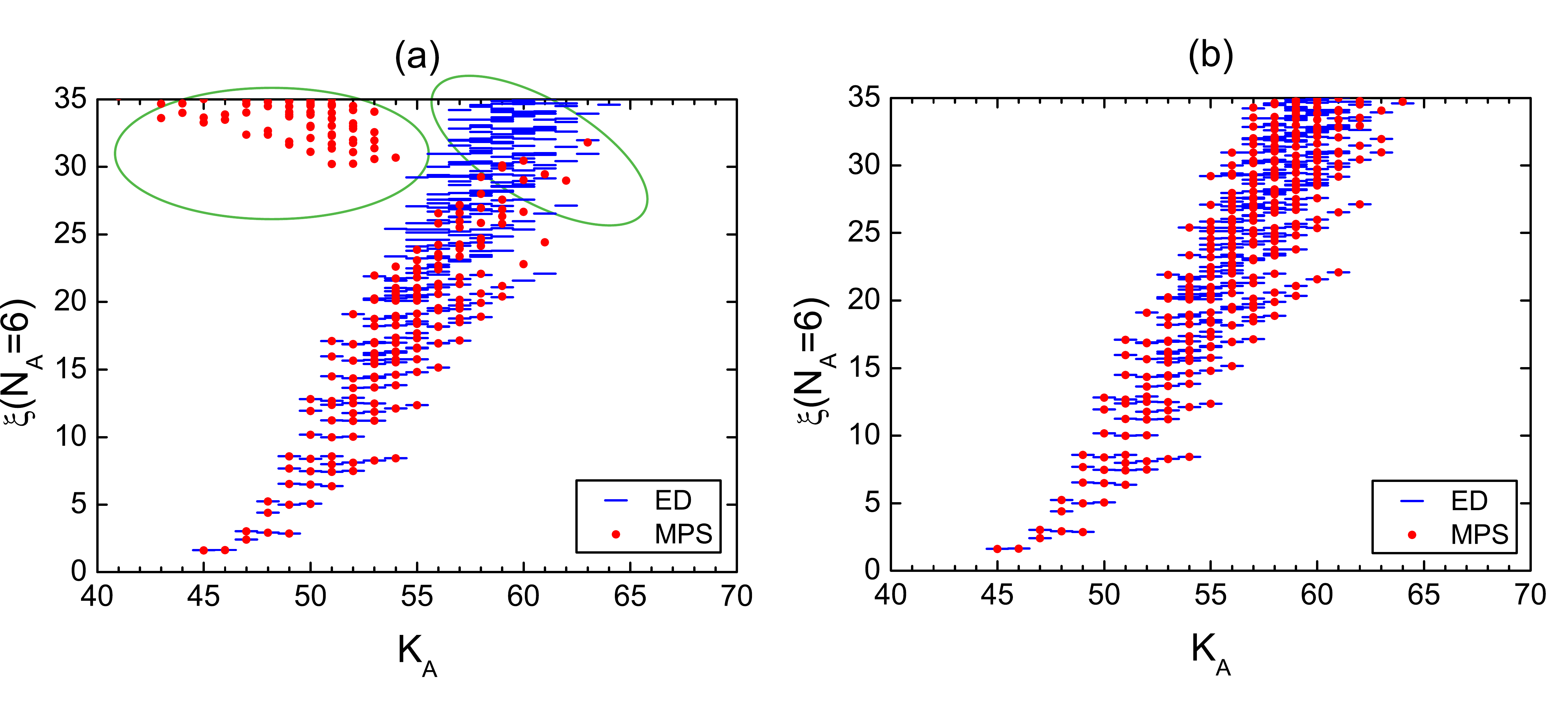}}
\caption{\label{mps_ed}The orbital-cut entanglement spectra obtained by ED (blue dashes) and MPS algorithm (red dots) for $N_e=12$ electrons at $\nu=1/3$ with $L=20$. $\Lambda=12,\varepsilon=10^{-6},\eta=10^{-10}$ in (a) and $\Lambda=12,\varepsilon=10^{-10},\eta=10^{-14}$ in (b). The difference from the ED results, which are caused by nonzero $\varepsilon$ and $\eta$, are indicated by the green circles in (a).}
\end{figure}
In Fig.~\ref{mps_ed}, we use the orbital-cut entanglement spectrum \cite{es}, which is a fingerprint of the topological order in FQH states, to analyze the error sources of our MPS algorithm. We observe two kinds of difference between the entanglement spectra obtained by ED and MPS algorithm, as indicated by the green circles in Fig.~\ref{mps_ed}(a). The extra levels in the left circle are caused by the relatively poor quality (large $\varepsilon$) of the
MPO representation of the Hamiltonian, and the missing levels in the right circle are
caused by the relatively large density matrix truncation error $\eta$. After improving the
quality of the MPO representation (reduce $\varepsilon$) and increase the accuracy of the density
matrix truncation (reduce $\eta$), we find the difference between MPS and ED results essentially
disappears, as shown in Fig.~\ref{mps_ed}(b).

\section{Performance of our MPS code}
We now apply our MPS code to systems at $\nu=1/3$ with sizes beyond the ED limit. Again, we choose the interaction between electrons as $\mathcal{V}_1$ pseudopotential, so the ground state in the $(N_e,K=3N_e(N_e-1)/2)$ sector is the exact Laughlin state with exactly zero ground energy.

\begin{figure}
\centerline{\includegraphics[width=0.5\linewidth]{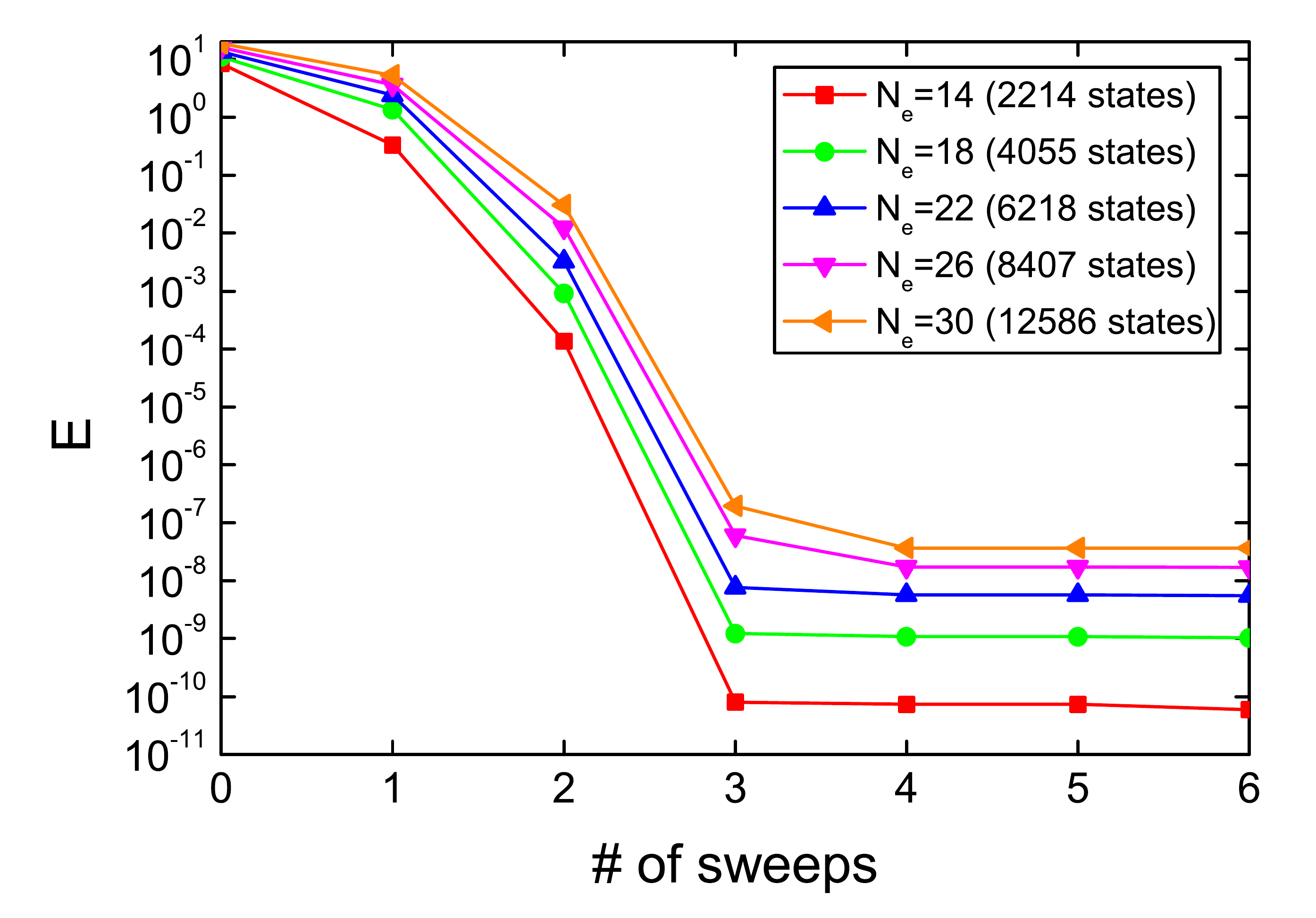}}
\caption{\label{mps}The ground energy at $\nu=1/3$ with $\Lambda=12,\varepsilon=10^{-8},\eta=10^{-13}$ versus the number of sweeps obtained our MPS algorithm for various system sizes. We consider square samples, namely the circumference of the cylinder $L=\sqrt{2\pi N_s}$. The maximal number of kept states is given in the bracket for each system size.
}
\end{figure}

In Fig.~\ref{mps}, we fix the accuracy of the MPO ($\Lambda$ and $\varepsilon$) and the density matrix truncation ($\eta$), and study the convergence of the ground energy on various square samples with $L=\sqrt{2\pi N_s}$. We choose the root configuration at $\nu=1/3$ \cite{root} as the initial state $|\Psi_0\rangle$. The number of kept states in the density matrix truncation is controlled by $\eta$ and changes during the sweeps. This is a little different from the usual DMRG algorithm where a fixed number of kept states is usually set at the beginning. However, in our MPS code, we can still track the number of kept states in each sweep and select the maximal one, which is also shown in Fig.~\ref{mps}. With the increase of the system size from $14$ electrons to $30$ electrons, the entanglement in the system grows due to the increase of circumference $L$. Thus the computational cost of the MPS simulation also increases, reflected by the fact that the maximal number of kept states grows by a factor of $6$. For the largest system size ($30$ electrons), our computation takes $19$ days by $15$ CPU cores on a computer cluster with $512$GB memory. The final energy that we obtain is very close to the theoretical value $0$ for each system size, but grows from roughly $10^{-11}$ for $14$ electrons to $10^{-8}$ for $30$ electrons. This is because higher accuracy is needed for larger system size and cylinder circumference.

\begin{figure}
\centerline{\includegraphics[width=0.5\linewidth]{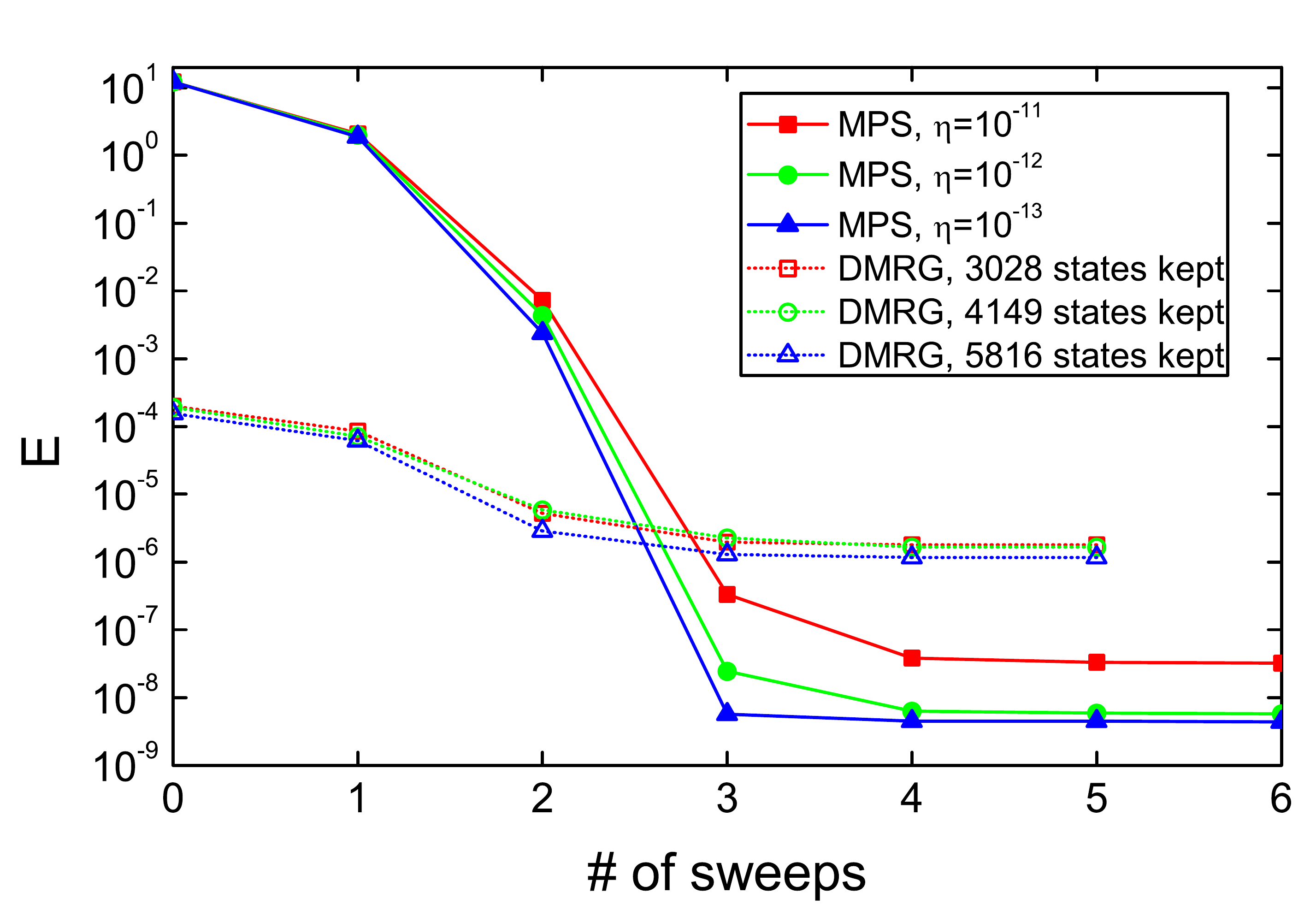}}
\caption{\label{mps_dmrg}The ground energy of $N_e=20$ electrons at $\nu=1/3$ with $L=20,\Lambda=12,\varepsilon=10^{-8}$ versus the number of sweeps for our MPS algorithm and traditional two-site DMRG algorithm. The number of kept states in the DMRG algorithm is set to be equal with the maximal number of kept states in the MPS sweeps.
}
\end{figure}

We also want to compare our MPS code with the traditional two-site DMRG code. We consider $20$ electrons at $\nu=1/3$ with $\mathcal{V}_1$ interaction and study the convergence of the ground energy for different density matrix truncations $\eta$ (Fig.~\ref{mps_dmrg}).
The number of kept states in the two-site DMRG algorithm is set to be equal with the maximal number of kept states in
the MPS sweeps. With the decrease of $\eta$, the ground energy for both of the MPS algorithm and two-site DMRG algorithm goes
closer to $0$. However, the MPS algorithm can reach lower energy than two-site DMRG.

\begin{figure}
\centerline{\includegraphics[width=0.5\linewidth]{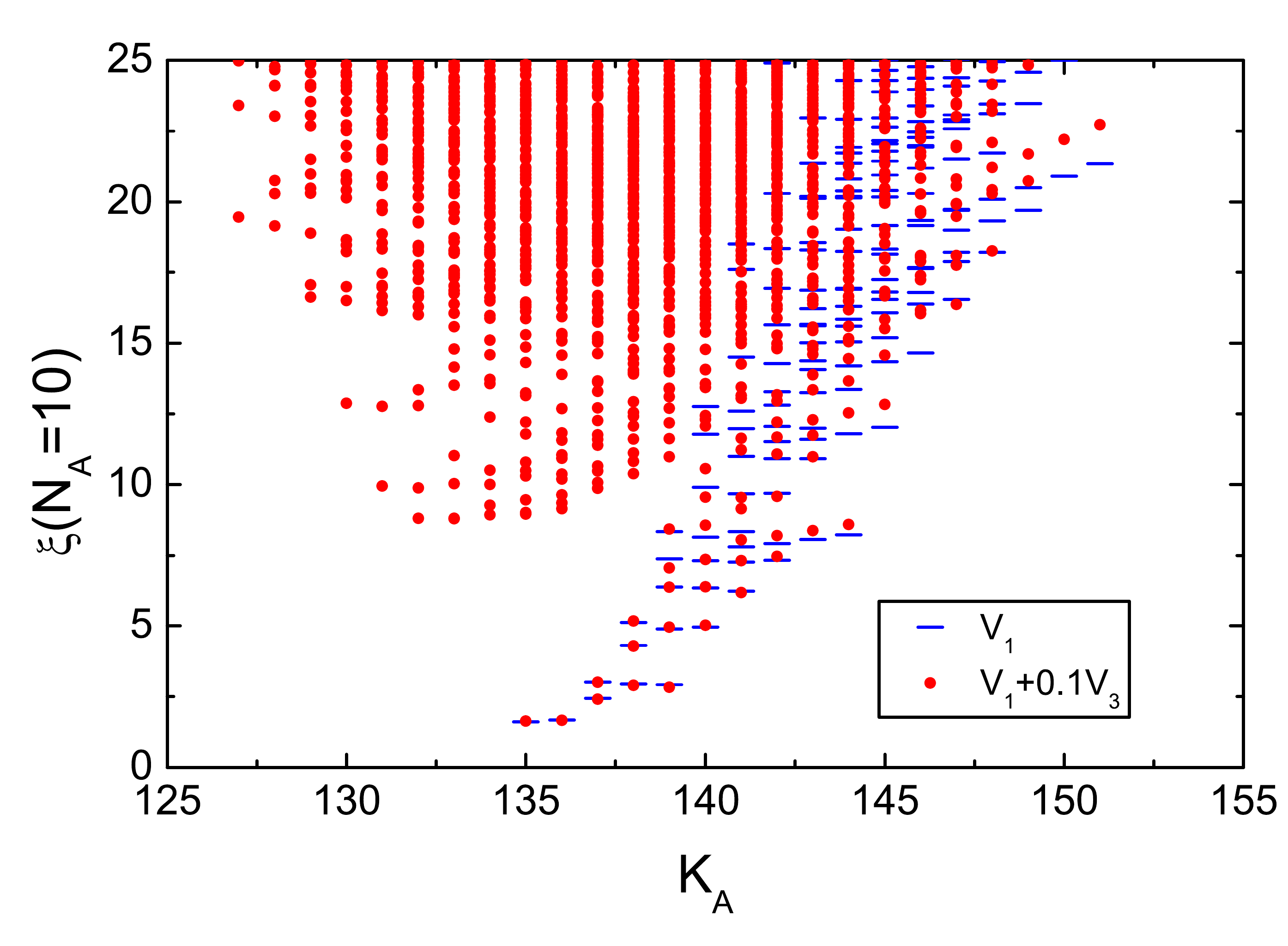}}
\caption{\label{v1_v3}The orbital-cut entanglement spectra obtained by MPS algorithm for pure $\mathcal{V}_1$ (green dashes) and $\mathcal{V}_1+0.1\mathcal{V}_3$ interactions (red dots) for $N_e=20$ electrons at $\nu=1/3$ with $L=20,\Lambda=12,\varepsilon=10^{-8},\eta=10^{-12}$.
}
\end{figure}

%\section{Go beyond the exact Laughlin state}
When the interaction goes beyond the pure $\mathcal{V}_1$ pseudopotential, the ground state at $\nu=1/3$ is no longer the exact Laughlin state with exactly zero ground energy. To achieve this, we use a combination of Haldane's $\mathcal{V}_1$ and $\mathcal{V}_3$ pseudopotentials with $\mathcal{V}_3=0.1\mathcal{V}_1$. Considering the strength of $\mathcal{V}_3$ is much smaller than $\mathcal{V}_1$, we expect that the ground state at $\nu=1/3$ is still in the Laughlin phase, although not the exact Laughlin state. Because $\mathcal{V}_3$ pseudopotential is a longer-range interaction than $\mathcal{V}_1$ pseudopotential, the bond dimension of the MPO is larger than that of pure $\mathcal{V}_1$ pseudopotential. We calculate the ground-state entanglement spectrum of $20$ electrons at filling $\nu=1/3$, and find that the low-lying part indeed matches that of the exact Laughlin state, although generic levels appear in the high energy region (Fig.~\ref{v1_v3}).

\section{Discussion}
In this work, we here reported a MPS code for finite FQH systems on the cylinder geometry. By comparing with ED and traditional two-site DMRG, we show its capability of searching for the FQH ground states. Compared with the MPS code for infinite FQH systems \cite{frank}, our code is more suitable to study the physics in finite systems, such as edge effects.

There are several possible directions for the future work. We can generalize our code to the infinite cylinder, as in Ref.~\cite{frank}, where a multi-site
update was used. It would be interesting to see the performance of single-site update and
density matrix correction in the that case.
We can also go beyond short-range Haldane's pseudopotential to deal with some long-range
Hamiltonians, such as dipole and Coulomb interactions. However, the bond
dimension of MPO increases fast with the interaction range. Therefore, we will need to
truncate the interaction differently and study results as a function of the truncation
length. Finally, because the only system-dependent part of our code is the MPO representation of
the physical Hamiltonian, it can readily be used in other many-body systems such as
spin chains and lattice models (such as fractional Chern insulators) \cite{fcidmrg1,fcidmrg2}.

\ack{We thank M.~P.~Zaletel, M.~C.~Banuls, and F.~D.~M.~Haldane for the useful discussion, and Sonika Johri for the DMRG calculation shown in Figure \ref{mps_dmrg}. This work was supported by the Department of Energy, Office of Basic Energy Sciences through Grant No.~DE-SC0002140.}

\end{document}